\documentclass[preprint,3p]{elsarticle} 
\usepackage{multicol,caption}

\newenvironment{Figure}
 {\par\medskip\noindent\minipage{\linewidth}}
 {\endminipage\par\medskip}

\usepackage{amssymb}

\usepackage{lineno}

\usepackage{graphicx}
\usepackage{dcolumn}
\usepackage{bm}
\usepackage{epsfig}
\usepackage{rotating}
\usepackage{float}
\usepackage{amsmath}
\usepackage{color}

\journal{Physics Letters B}

\begin{document}

\begin{frontmatter}



\title{Measurement of the leptonic decay width of $J/\psi$ using initial state radiation} 

\begin{small}
\author{
M.~Ablikim$^{1}$, M.~N.~Achasov$^{9,f}$, X.~C.~Ai$^{1}$, O.~Albayrak$^{5}$, M.~Albrecht$^{4}$, D.~J.~Ambrose$^{44}$, A.~Amoroso$^{48A,48C}$, F.~F.~An$^{1}$, Q.~An$^{45,a}$, J.~Z.~Bai$^{1}$
% ...
\\
\vspace{0.2cm}
(BESIII Collaboration)\\
\vspace{0.2cm} {\it
$^{1}$ Institute of High Energy Physics, Beijing 100049, People's Republic of China\\
% ...
$^{4}$ Bochum Ruhr-University, D-44780 Bochum, Germany\\
$^{5}$ Carnegie Mellon University, Pittsburgh, Pennsylvania 15213, USA\\
% ...
$^{9}$ G.I. Budker Institute of Nuclear Physics SB RAS (BINP), Novosibirsk 630090, Russia\\
% ...
$^{44}$ University of Rochester, Rochester, New York 14627, USA\\
$^{45}$ University of Science and Technology of China, Hefei 230026, People's Republic of China\\
% ...
$^{48}$ (A)University of Turin, I-10125, Turin, Italy; (B)University of Eastern Piedmont, I-15121, Alessandria, Italy; (C)INFN, I-10125, Turin, Italy\\
% ...
$^{a}$ Also at State Key Laboratory of Particle Detection and Electronics, Beijing 100049, Hefei 230026, People's Republic of China\\
% ...
$^{f}$ Also at the Novosibirsk State University, Novosibirsk, 630090, Russia\\
% ...
}}

\vspace{0.4cm}
\end{small}



\begin{abstract}
Using a data set of 2.93 fb$^{-1}$ taken at a center-of-mass energy of $\sqrt{s}$ =  3.773 GeV  with the BESIII detector at the BEPCII collider, we measure the process $e^+e^-\rightarrow J/\psi\gamma\rightarrow \mu^+\mu^-\gamma$ and determine the product of the branching fraction and the electronic width $\mathcal B_{\mu\mu}\cdot \Gamma_{ee} = (333.4 \pm 2.5_{\rm stat} \pm 4.4_{\rm sys})$~eV. Using the earlier-published BESIII result for $\mathcal B_{\mu\mu}$  = (5.973 $\pm$ 0.007$_{\rm stat}$ $\pm$ 0.037$_{\rm sys}$)\%, we derive the $J/\psi$ electronic width $\Gamma_{ee}$~= (5.58 $\pm$ 0.05$_{\rm stat}$ $\pm$ 0.08$_{\rm sys}$) keV. 
\end{abstract}

\end{frontmatter}



\begin{multicols}{2}


The electronic width of the $J/\psi$ resonance \mbox{$\Gamma_{ee}\equiv\Gamma_{ee}(J/\psi)$} has been measured by BaBar~\cite{JPsi_babar} and CLEO-c~\cite{JPsi_CLEO}, employing the technique of Initial State Radiation (ISR), in which one of the
beam particles radiates a photon. Consequently, the invariant mass range below the center-of-mass energy of the $e^+e^-$ collider becomes available. Using a different method, the {\sc kedr} experiment also measured its electronic width with improved precision~\cite{JPsi_KEDR}. In this paper, we study the process $e^+e^-\rightarrow \mu^+\mu^-\gamma$ using the ISR method with $\mu^+\mu^-$ invariant mass $m_{2\mu}$ between 2.8 and 3.4 GeV/$c^2$, which covers the charmonium resonance $J/\psi$. The cross section $\sigma_{J/\psi\gamma} \equiv \sigma(e^+e^-\rightarrow J/\psi \gamma\rightarrow \mu^+\mu^-\gamma)$ is proportional to $\Gamma_{ee}\cdot\mathcal B_{\mu\mu}$, where \mbox{$\mathcal B_{\mu\mu} \equiv \mathcal B(J/\psi\rightarrow\mu^+\mu^-)$} is the branching fraction of the muonic decay of the $J/\psi$ resonance. With the precise measurement of $\mathcal B_{\mu\mu}$ from BESIII~\cite{branching_BES}, we have the opportunity to obtain $\Gamma_{ee}$ with high precision.
The differential cross section of $\sigma_{J/\psi \gamma}$ can be expressed in terms of the center-of-mass energy squared $s$ as
\begin{linenomath}
\begin{equation}
	\frac{d\sigma_{J/\psi}(s,m_{2\mu})}{dm_{2\mu}} = \frac{2m_{2\mu}}{s}W(s,m_{2\mu})  BW(m_{2\mu}),
	\label{JPsi_crossSection}
\end{equation}
\end{linenomath}
where $W(s,m_{2\mu})$ is the radiator function, describing the probability that one of the beam particles emits an ISR photon~\cite{radiator function}, and $BW(m_{2\mu})$ is the Breit-Wigner function. $W(s,m_{2\mu})$ is calculated by the {\sc phokhara} event generator, with an estimated accuracy of 0.5\%~\cite{Phokhara}. The Breit-Wigner function is 
\begin{linenomath}
\begin{linenomath}
\begin{equation}
	BW(m_{2\mu}) =  \frac{12\pi\mathcal B_{\mu\mu}\cdot\Gamma_{ee}\Gamma_{\rm tot}}{(m_{2\mu}^2 - M_{J/\psi}^2)^2 + M_{J/\psi}^2\Gamma_{\rm tot}^2}	 ,
\end{equation}
\end{linenomath}
\end{linenomath}
\cite{PDG2014} in which $\Gamma_{\rm tot}$ and $M_{J/\psi}$ are the $J/\psi$ full width and mass. Both values are taken from the world averages~\cite{PDG2014}. The cross section $\sigma_{J/\psi \gamma}$ over a specified $m_{2\mu}$ range can be expressed using:
\begin{linenomath}
\begin{equation}
	\sigma_{J/\psi \gamma} (s) = \frac{N_{J/\psi}}{\epsilon\cdot\mathcal L} = \Gamma_{ee}\cdot \mathcal B_{\mu\mu}\cdot I(s),
	\label{formula_crossSection}
\end{equation}
\end{linenomath}
where $N_{J/\psi}$ is the number of signal events within the mass range after background subtraction, $\epsilon$ is the selection efficiency obtained from a Monte Carlo (MC) simulation, $\mathcal L$ is the integrated luminosity of the data set, and $I(s)$ is the integral
\begin{linenomath}
\begin{equation}
\label{formula_I}
		I(s) \equiv \int_{m_\text{min}}^{m_\text{max}} \frac{2m_{2\mu}}{s}W(s,m_{2\mu}) b(m_{2\mu})  dm_{2\mu},
\end{equation}
\end{linenomath}
in which $b(m_{2\mu}) \equiv BW(m_{2\mu}) / \Gamma_{ee}\cdot\mathcal B_{\mu\mu}$. A mass range between $m_\text{min} = 2.8$~GeV/$c^2$ and \mbox{$m_\text{max} = 3.4$~GeV/$c^2$} is chosen in which $N_{J/\psi}$ is determined.

The above equations do not take into account interference effects of the resonant $\mu^+\mu^-$ production via $J/\psi$ and the non-resonant $e^+e^-\rightarrow \mu^+\mu^-\gamma$ QED production. At lowest order in the fine structure constant $\alpha$, these can be included by replacing $BW(m_{2\mu})$ by~\cite{interference}
\begin{linenomath}
\begin{equation}
	BW'(m_{2\mu}) = \frac{4\pi\alpha^2}{3m_{2\mu}^2}\bigg( \big|1 - \zeta (m_{2\mu}) \big|^2 - 1\bigg)
	\label{formula_BW_theo},
\end{equation}
\end{linenomath}
with
\begin{linenomath}
\begin{equation}
	\zeta (m_{2\mu}) = \frac{3}{\alpha} \cdot\frac{\sqrt{\mathcal B_{\mu\mu}\cdot\Gamma_{ee}\Gamma_{\rm tot}}M_{J/\psi}}{M_{J/\psi}^2 - m_{2\mu}^2 - iM_{J/\psi}\Gamma_{\rm tot}}
\end{equation}
\end{linenomath}
and $b(m_{2\mu})$ by its equivalent $b'(m_{2\mu}) \equiv BW'(m_{2\mu}) / \Gamma_{ee}\cdot\mathcal B_{\mu\mu}$. The interference is non-symmetrical around the peak; destructive below and constructive above. The radiator function gives a larger weight to lower photon energies, corresponding to higher di-muon invariant masses. This changes the $m_{2\mu}$ shape around the peak asymmetrically. Replacing $b(m_{2\mu})$ by $b'(m_{2\mu})$ in formula (\ref{formula_I}) and using the world average~\cite{PDG2014} for $\Gamma_{ee}\cdot\mathcal B_{\mu\mu}$ enhances $I(s)$ by about 2.2\%. The function $b'(m_{2\mu})$ depends on $\Gamma_{ee}\cdot\mathcal B_{\mu\mu}$. Hence, an iterative procedure is used for its extraction.

We use $e^+e^-$ collision data collected at the Beijing Spectrometer III (BESIII) experiment.
The BESIII detector~\cite{BESIII} is located at the double-ring $e^+e^-$ Beijing Electron Positron Collider (BEPCII). The cylindrical BESIII detector covers 93\% of the full solid angle. It consists of the following detector systems: (1) A Multilayer Drift Chamber (MDC) filled with a Helium-based gas, composed of 43 layers, providing a spatial resolution of 135 $\mu$m and a momentum resolution of 0.5\% for charged tracks at \mbox{1 GeV/$c$} in a magnetic field of 1 T.
\mbox{(2) A} Time-of-Flight system (TOF),  composed of 176 plastic scintillator counters in the barrel part, and 96 counters in the endcaps. The time resolution in the barrel is 80 ps and 110 ps in the endcaps. For momenta up to 1 GeV/$c$ a 2$\sigma$ K/$\pi$ separation is obtained.
(3) A CsI(Tl) Electro-Magnetic Calorimeter (EMC), with an energy resolution of 2.5\% in the barrel and 5\% in the endcaps at an energy of 1 GeV.
(4) A Muon Chamber (MUC) consisting of nine barrel and eight endcap resistive plate chamber layers with a 2 cm position resolution.  \\



We analyze 2.93 fb$^{-1}$~\cite{2pi_BES} of data taken at \mbox{$\sqrt{s} = 3.77$3~GeV} in two separate runs in 2010 and 2011. A  {\sc Geant4}-based~\cite{GEANT1,GEANT2} Monte Carlo (MC) simulation is used to determine efficiencies and study backgrounds.  To simulate the ISR process $e^+e^-\rightarrow \mu^+\mu^-\gamma$, we use the {\sc phokhara} event generator~\cite{Phokhara,Phokhara7}. It includes ISR and final state radiation (FSR) corrections up to next-to-leading order (NLO). Hadronic ISR production is also simulated with {\sc phokhara}. Bhabha scattering is simulated using the {\sc babayaga} 3.5 event generator~\cite{BABAYAGA}.  Continuum MC is produced with the {\sc kkmc} generator~\cite{KKMC}.
\\

We require the presence of at least two charged tracks in the MDC with net charge zero. The points of closest approach from the interaction point (IP) for these two tracks are required to be within a cylinder of 1 cm radius in the transverse direction and $\pm10$~cm of length along the beam axis. In case of three-track events, we choose the track pair with net charge zero which is closest to the IP.
The polar angle $\theta$ of the tracks is required to be found in the fiducial volume of the MDC, $0.4\;\text{rad}< \theta < \pi-0.4\;\text{rad}$, where $\theta$ is the polar angle of the track with respect to the beam axis. We require the transverse momentum $p_t$ to be greater than 300 MeV/$c$ for each track.
To enhance statistics and to suppress non-ISR background, we investigate untagged ISR events, where the ISR photon is emitted under a small angle $\theta_\gamma$, almost collinear with the beam, and therefore does not end up in the fiducial volume of the EMC. This is a new approach with respect to BaBar and CLEO-c (both used tagged ISR photons), which has been proved to be valid and effective by using the {\sc phokhara} event generator~\cite{phokhara-add}. A one-constraint (1C) kinematic fit is performed under the hypothesis $e^+e^-\rightarrow\mu^+\mu^-\gamma$, using as input the two selected charged track candidates as well as the four-momentum of the initial $e^+e^-$ system. The constraint is a missing massless particle. The fit imposes overall energy and momentum balance. The $\chi^2_{1C}$ value returned by the fit is required to be smaller than 10. In addition, the predicted missing photon angle with respect to the beam axis, $\theta_\gamma$, has to be smaller than $0.3$ radians or greater than $\pi - 0.3$ radians in the lab frame.
Radiative Bhabha scattering $e^+e^-\gamma(\gamma)$ has a cross section that is up to three orders of magnitude larger than the signal cross section. Therefore, electron tracks need to be suppressed. An electron particle identification (PID) algorithm is used for this purpose, employing information from the MDC, TOF and EMC~\cite{BESIII_2}. The probabilities for the track being a muon $P(\mu)$ or an electron $P(e)$ are calculated, and $P(\mu)>P(e)$ is required for both charged tracks, which leads to
an electron suppression of more than 96\%. To further suppress hadronic background, an Artificial Neural Network (ANN) built on the TMVA package~\cite{TMVA} is used. The ANN is described in detail in Ref.~\cite{2pi_BES}. Both charged tracks are required to have a classifier output value $y_{\rm ANN}$ of this method smaller than 0.3 to be treated as muons, leading to a signal loss of less than 30\% and a background rejection of more than 99\%.\\

Background beyond the radiative processes $\mu^+\mu^-\gamma$ is studied with MC simulations. Table \ref{JPsi_background} lists the number of events remaining after all previously described requirements in the mass range between 2.8 and 3.4 GeV/$c^2$. About $4.8 \times 10^5$ events are found in the data within this range. The background fraction is found to be smaller than 0.04\% for each of the 150 $m_{2\mu}$ mass bins. We subtract it from the data bin by bin. \\
\begin{Figure}
	\centering
	\captionof{table}{Total number of non-muon background events between 2.8 $\le$ $m_{2\mu}$ $\le$ 3.4 GeV/$c^2$ obtained with MC samples, which are normalized to the luminosity of the data set.}
	\begin{tabular}{ l | c } 
	\hline \hline
	\bf{Final state}		&	\bf{Background events} 	\\  \hline
	$e^+e^-(\gamma)$				& negl.	\\
	$\pi^+\pi^-\gamma$				& $8.4\pm 2.9$\\	
	$\pi^+\pi^-\pi^0\gamma$			& $3.3\pm 1.8$\\
	$\pi^+\pi^-\pi^0\pi^0\gamma$		& $0.3\pm 0.6$	\\
	$\pi^+\pi^-\pi^+\pi^-\gamma$		& negl.	\\
	$K^+K^-\gamma$				& $1.7\pm 1.3$\\
	$K^0\overline{ K^0}\gamma$		& negl.	\\
	$p\overline{p}\gamma$			& negl.	\\
	Continuum					& $1.7 \pm 1.3$ \\
	$\psi(3770)\rightarrow D^+D^-$					& negl.	\\
	$\psi(3770)\rightarrow D^0\overline{D^0}$			& negl.	\\
	$\psi(3770)\rightarrow \rm{non} \; D\overline{D}$			& $11.2\pm3.4$\\
	$J/\psi\rightarrow \rm{non} \; \mu\mu$			& $11.8\pm3.5$ 	\\
\hline \hline
	\end{tabular}
	\label{JPsi_background}
\end{Figure}

The selection efficiency $\epsilon$ is determined based on signal MC events. It is obtained as the ratio of the measured number of events after all selection requirements $N^{\rm true}_{\rm measured}$ to all generated ones $N^{\rm true}_{\rm generated}$ only. The true MC sample of $J/\psi$ decays with the full $\theta_\gamma$ range, which does not contain the detector reconstruction, is used here by applying efficiency corrections to each track for muon tracking reconstruction, electron-PID, and ANN efficiency. These corrections have been derived in Ref.~\cite{2pi_BES}. We find $\epsilon$ to be \mbox{(32.04 $\pm$ 0.09)\%}, where the error is due to the size of the signal MC sample. 
\\

\begin{figure*}
   \centering
   \includegraphics[width=6.5cm]{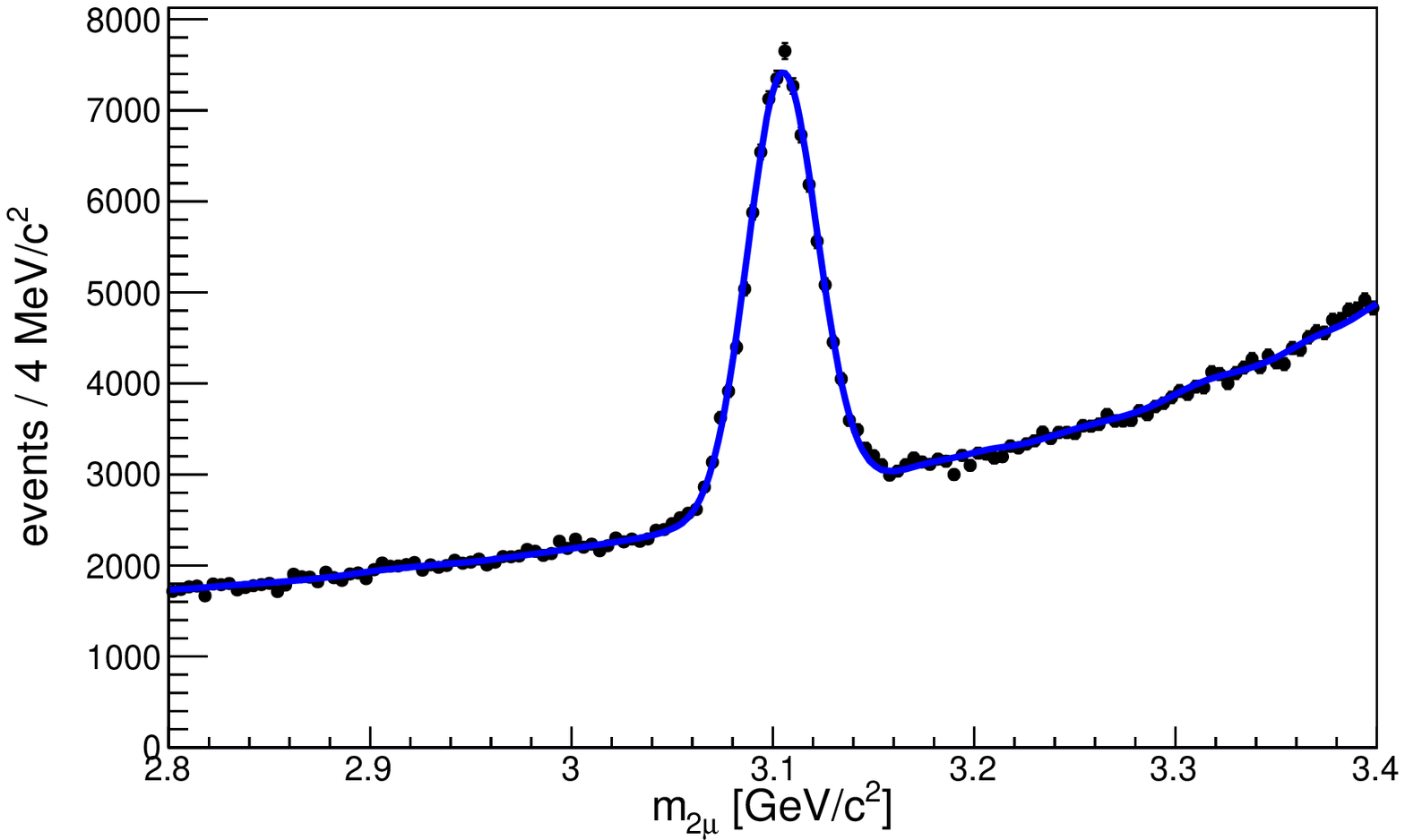}
     \captionof{figure}{Fit to the data using the final value of $\Gamma_{ee}\cdot\mathcal B_{\mu\mu}$ from Table \ref{table_iteration} in the MC histogram for the fit.}
   \label{fit}
\end{figure*}

The number of $J/\psi$ events $N_{J/\psi}$ is determined from a binned maximum likelihood fit to data. The fit function $f(x)$ used is
\begin{linenomath}
\begin{equation}
	f(x) = N_{J/\psi} \big[M(x) \otimes G(x)\big] + \big(N_\text{total} - N_{J/\psi}\big) p(x),
	\label{formula_fit}
\end{equation}
\end{linenomath}
where $M(x)$ describes the shape of the MC-simulated $J/\psi$ peak. We extract the shape from a MC simulation of the $J/\psi$ production using a certain $\Gamma_{ee}\cdot\mathcal B_{\mu\mu}$ value as an input, together with QED $\mu^+\mu^-\gamma$ production (including interference effects) as simulated with the {\sc phokhara} event generator. Then, the histogram $M(x)$ is obtained by subtracting a pure QED $\mu^+\mu^-\gamma$ MC sample. It is shown in Fig.~\ref{MC_shape}, using the world average~\cite{PDG2014} for $\Gamma_{ee}\cdot\mathcal B_{\mu\mu}$ as input.
To take into account differences in mass resolutions between data and MC simulation, $M(x)$ is convolved (denoted by the operator $\otimes$) with a Gaussian distribution $G(x)$ with mean $\bar x$ and width $\sigma$, whose parameters are determined by the fit to data.
To describe the non-resonant QED production in the fit, a polynomial of fourth order is used,
\begin{linenomath}
\begin{equation}
	p(x) = \sum_{i=0}^4 a_i x^{i} \; .
\end{equation}
\end{linenomath}
$N_{\rm total}$ is the constant number of data events between 2.8 and 3.4 GeV/$c^2$. Free parameters in the fit are $N_{J/\psi}$, $\bar x$, $\sigma$, and the coefficients $a_i$ \mbox{(i = 1,...,4)}. Hence, $N_{J/\psi}$ can be obtained directly by the fit. The fit result is shown in Fig.~\ref{fit}; we find $\bar x = (2.6 \pm 0.1)\;\text{MeV}/c^2$, $\sigma = (10.5 \pm 0.2)\;\text{MeV}/c^2$, and $\chi^2/ndf = 149.8/143$.

Equation (\ref{formula_crossSection}) is used to determine $\Gamma_{ee}\cdot\mathcal B_{\mu\mu}$ in an iterative process. In each iteration, we simulate the histogram $M(x)$ and calculate $I(s)$ (including interference corrections), using a $\Gamma_{ee}\cdot\mathcal B_{\mu\mu}$ input value, and extract the $\Gamma_{ee}\cdot\mathcal B_{\mu\mu}$ output with Eq.~(\ref{formula_crossSection}). This result is used as input for the next iteration. We choose the PDG value~\cite{PDG2014} as the starting value. The results of each iteration are summarized in Table \ref{table_iteration}. After three iterations the result becomes stable within four decimal places, which corresponds to the experimental uncertainty. As the final value we find 
\begin{linenomath}
\begin{equation*}
\Gamma_{ee}\cdot\mathcal B_{\mu\mu} = (333.4 \pm 2.5_{\rm stat}  \pm 4.4_{\rm sys})\;\rm eV,
\end{equation*}
\end{linenomath}
where the first error is the statistical uncertainty from the fit procedure, and the second error is the systematic uncertainty. 

\begin{Figure}
   \centering
   \includegraphics[width=6.5cm]{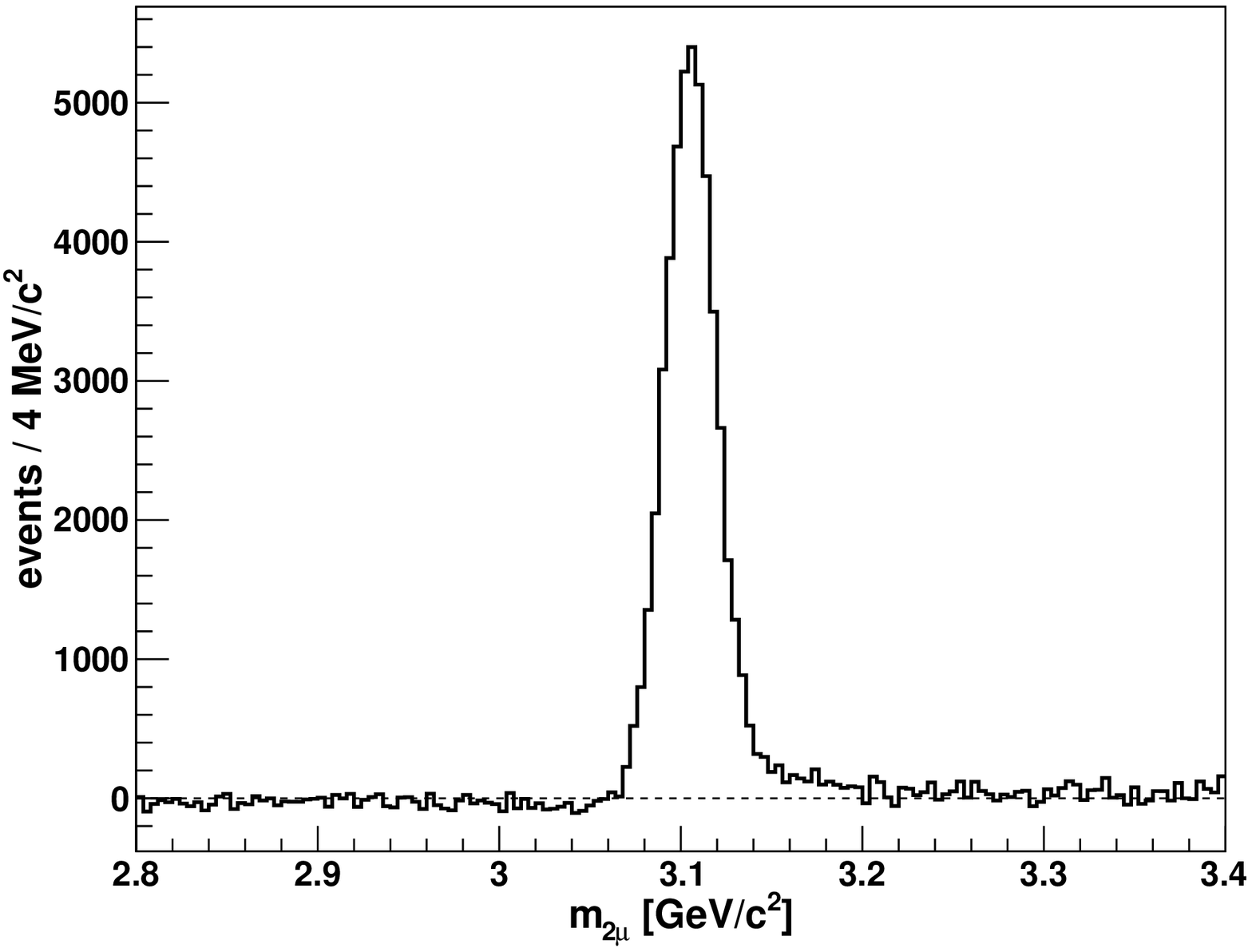}
   \captionof{figure}{MC histogram from the {\sc phokhara} generator after full detector simulation used for the fit. The value of $\Gamma_{ee}\cdot\mathcal B_{\mu\mu}$ used for generation is the one from Ref.~\cite{PDG2014}.}
   \label{MC_shape}
\end{Figure}

All systematic uncertainties are summarized in Table \ref{JPsi_uncertainties}. They are summed up in quadrature to be 1.3\%. They are derived as follows:\\
\noindent (1) Integral $I(s)$: The difference in $I(s)$, when enhancing or decreasing the value of $\Gamma_{ee}\cdot\mathcal B_{\mu\mu}$ within five standard deviations of the error, claimed by Ref.~\cite{PDG2014}, is smaller than 0.2\%. This deviation is considered as the systematic uncertainty of accommodating the interference effects in $I(s)$. \\
\noindent (2) Background subtraction: A conservative uncertainty of 100\% is assumed for the MC samples. Hence, the systematic uncertainty due to background subtraction is smaller than 0.04\% per bin and can therefore be neglected.\\
\noindent (3) Efficiency $\epsilon$: The data-MC efficiency corrections have been studied in Ref.~\cite{2pi_BES}. The corresponding systematic uncertainties are listed in Table~\ref{JPsi_uncertainties}. They are found to be smaller than 0.5\% in each case.\\
\noindent (4) To estimate the uncertainty introduced by the requirements on $\theta_{\gamma}$ and $\chi^2_{1C}$, the resolution differences between data and MC simulation in these variables are obtained. In case of $\theta_\gamma$, we find the resolution difference to be $(66 \pm 3) \times 10^{-5}$ radians, by comparing an ISR photon tagged clean $\mu^+\mu^-\gamma$ sample both from data and MC simulation. In case of $\chi^2_{1C}$, we determine the efficiency of the applied requirement $\chi^2_{1C} < 10$ in data and MC simulation. We vary this requirement in data such that the efficiencies in data and MC simulation are the same. The difference to the actually used requirement is taken as resolution difference, which we find to be $(1.1 \pm 0.1)$ units in $\chi^2_{1C}$. To achieve a better description of $\epsilon$, both variables are smeared in the signal MC sample with a Gaussian with a mean value of zero and a width corresponding to the resolution difference. To estimate the contribution to the systematic uncertainty, these variables are also varied with a $\pm 1$ standard deviation, and the difference in $\epsilon$ is taken as the systematic uncertainty, which is found to be less than 0.5\% for $\chi^2_{1C}$ and negligible for $\theta_\gamma$.\\
\noindent (5) The chosen mass range between 2.8 and 3.4 GeV/$c^2$ is varied within 0.1 GeV/$c^2$, using the final value of $\Gamma_{ee}\cdot \mathcal B_{\mu\mu}$ after the iteration procedure. The difference in $\Gamma_{ee}\cdot \mathcal B_{\mu\mu}$ is smaller than 0.3\%, and is used as a systematic uncertainty.\\
\noindent (6) The luminosity has been measured in Refs.~\cite{lumi_BES,2pi_BES} with an uncertainty of 0.5\%.\\
\noindent (7) The radiator function is extracted from the {\sc phokhara} event generator~\cite{Phokhara7} and has an uncertainty of 0.5\%. \\
\noindent (8) The angular acceptance of the charged tracks is studied by varying this requirement by more than three standard deviations of the angular resolution, and studying the corresponding difference in the final result. An uncertainty of less than 0.1\% is found.  \\

\begin{Figure}
	\centering
	\captionof{table}{Summary of the systematic uncertainties.}

	\begin{tabular}{ l  c } 
	\hline \hline 
	\bf{Source}		&	\bf{Uncertainty}	\\
					&	\bf{(\%)}	\\	
	\hline
	Background subtraction		&	negl.		\\
	Muon tracking efficiency		&	0.5		\\	
	Muon ANN efficiency		&	0.5		\\	
	Muon e-PID efficiency		&	0.5		\\	
	1C kinematic fit				& 	0.5 		\\
	Angular acceptance			&	0.1 		\\
	Luminosity				&	0.5 		\\
	Radiator function 			&	0.5		\\
	Parametrizing the interference 	& 0.2		\\
	Variation of fit range		&	0.3		\\
	\hline
	\bf{Sum}					&	1.3	\\
	\hline \hline
	
	\end{tabular}
	\label{JPsi_uncertainties}
\end{Figure}

With \mbox{$\mathcal B_{\mu\mu}$ = (5.973 $\pm$ 0.007$_{\rm stat}$ $\pm$ 0.038$_{\rm sys}$)\%} from an independent BESIII measurement~\cite{branching_BES}, our measurement yields
\begin{linenomath}
\begin{equation*}
	\Gamma_{ee} = (5.58 \pm 0.05_{\rm stat} \pm 0.08_{\rm sys}) \, \rm keV.
\end{equation*}
\end{linenomath}
Our measurement of  $\Gamma_{ee}\cdot \mathcal B_{\mu\mu}$ is consistent with the results from BaBar~\cite{JPsi_babar},  CLEO-c~\cite{JPsi_CLEO} and KEDR~\cite{JPsi_KEDR}. The measured value for $\Gamma_{ee}$ is more precise, as summarized in Table \ref{JPsi_comparison}.

\begin{Figure}
	\centering
	\captionof{table}{Results of the iteration steps. As the starting value, the PDG 2014 one is used. The errors are the statistical ones.}
	\begin{tabular}{ c  c   c} 
	\hline \hline 
	\bf{Step}		&	\bf{$\Gamma_{ee}\cdot \mathcal B_{\mu\mu}$}	&	\bf{$\Gamma_{ee}\cdot \mathcal B_{\mu\mu}$}\\
	&	\bf{input value}	&	\bf{output value [eV]}\\
	\hline
	1	&	PDG value~\cite{PDG2014}	&	$333.9 \pm 2.5$ \\
	2	&	result of step 1		&	$333.3 \pm 2.5$\\
	3	&	result of step 2		&	$333.4 \pm 2.5$\\	
	4	&	result of step 3		&	$333.4 \pm 2.5$\\
	\hline \hline	
	\end{tabular}
	\label{table_iteration}
\end{Figure}

In summary, we have used the ISR process $e^+e^-\rightarrow J/\psi \gamma\rightarrow \mu^+\mu^-\gamma$ to measure \mbox{$\Gamma_{ee}\cdot\mathcal B_{\mu\mu}$ = (333.4 $\pm$ 2.5$_{\rm stat}$  $\pm$ 4.4$_{\rm sys}$) eV} with a total relative uncertainty of 1.5\%. Combined with the BESIII measurement of $B_{\mu\mu}$, we obtain \mbox{$\Gamma_{ee} = (5.58 \pm 0.05_{\rm stat} \pm 0.08_{\rm sys})$ keV} with a relative precision of 1.7\%.\\

\begin{figure*}
	\centering
	\captionof{table}{Results of the BaBar~\cite{JPsi_babar}, CLEO-c~\cite{JPsi_CLEO} and KEDR~\cite{JPsi_KEDR} measurements compared to this work.}
	\begin{tabular}{ l  l  l  l} 
	\hline \hline 
	\bf{Measurement}	&	$\Gamma_{ee}\cdot\mathcal B_{\mu\mu}$ [eV]	&	Used $\mathcal B_{\mu\mu}$ value [\%]	&		$\Gamma_{ee}$ [keV]\\
	\hline
	BaBar		&	330.1 $\pm$ 7.7$_{\rm stat}$  $\pm$ 7.3$_{\rm sys}$	&	5.88 $\pm$ 0.10~\cite{branching_BABAR} &	5.61 $\pm$ 0.20\\
	
	CLEO-c		&	338.4 $\pm$ 5.8$_{\rm stat}$  $\pm$ 7.1$_{\rm sys}$	& 	5.953 $\pm$ 0.056$_{\rm stat}$ $\pm$ 0.042$_{\rm sys}$~\cite{branching_CLEO}	&	5.68 $\pm$ 0.11$_{\rm stat}$ $\pm$ 0.13$_{\rm sys}$\\
	
	KEDR               &      331.8 $\pm$ 5.2$_{\rm stat}$ $\pm$ 6.3$_{\rm sys}$         &      5.94 $\pm$ 0.06~\cite{PDG2008}  & 5.59 $\pm$ 0.12 \\
	
	\bf{This work}	&	333.4 $\pm$ 2.5$_{\rm stat}$  $\pm$ 4.4$_{\rm sys}$	& 	5.973 $\pm$ 0.007$_{\rm stat}$ $\pm$ 0.037$_{\rm sys}$~\cite{branching_BES}	&	5.58 $\pm$ 0.05$_{\rm stat}$ $\pm$ 0.08$_{\rm sys}$\\
	\hline \hline
	\end{tabular}
	\captionsetup{font=footnotesize}
	\label{JPsi_comparison}
\end{figure*}



The BESIII collaboration thanks the staff of BEPCII and the IHEP computing center for their strong support. This work is supported in part by National Key Basic Research Program of China under Contract No. 2015CB856700; National Natural Science Foundation of China (NSFC) under Contracts Nos. 11125525, 11235011, 11322544, 11335008, 11425524; the Chinese Academy of Sciences (CAS) Large-Scale Scientific Facility Program; the CAS Center for Excellence in Particle Physics (CCEPP); the Collaborative Innovation Center for Particles and Interactions (CICPI); Joint Large-Scale Scientific Facility Funds of the NSFC and CAS under Contracts Nos. 11179007, U1232201, U1332201; CAS under Contracts Nos. KJCX2-YW-N29, KJCX2-YW-N45; 100 Talents Program of CAS; INPAC and Shanghai Key Laboratory for Particle Physics and Cosmology; German Research Foundation DFG under Contract No. Collaborative Research Center CRC-1044; Istituto Nazionale di Fisica Nucleare, Italy; Ministry of Development of Turkey under Contract No. DPT2006K-120470; Russian Foundation for Basic Research under Contract No. 14-07-91152; U. S. Department of Energy under Contracts Nos. DE-FG02-04ER41291, DE-FG02-05ER41374, DE-FG02-94ER40823, DESC0010118; U.S. National Science Foundation; University of Groningen (RuG) and the Helmholtzzentrum fuer Schwerionenforschung GmbH (GSI), Darmstadt; WCU Program of National Research Foundation of Korea under Contract No. R32-2008-000-10155-0.





\end{multicols}
\end{document}